# A tale of two twins

L.Benguigui

Physics Department and Solid State Institute
Technion-Israel Institute of Technology
32000   Haifa   Israel

Abstract

The thought experiment (called the clock paradox or the twin paradox) proposed by Langevin in 1911 of two observers, one staying on earth and the other making a trip toward a Star with a velocity near the light velocity is very well known for its very surprising result. When the traveler comes back, he is younger than the stay on Earth. This astonishing situation deduced form the theory of Special Relativity   sparked a huge amount of articles, discussions and controversies such that it remains a particular phenomenon probably unique in Physics. In this article we propose to study it. First, we lookedl for the simplest solutions when one can observe that the published solutions correspond in fact to two different versions of the experiment. It appears that the complete and simple  solution of Møller is neglected for complicated  methods with dubious validity.  We propose to interpret this avalanche of works by the difficulty to accept the conclusions of the Special Relativity, in particular the difference in the times indicated by two clocks, one immobile and the second moving and finally stopping. We suggest also that the name "twin paradox" is maybe related to some subconscious idea concerning conflict between twins as it can be found in the Bible and in several mythologies.

# Introduction

The thought experiment in the theory of Relativity called the "twin paradox" or the "clock paradox" is very well known in physics and even by non-physicists. In consist in imagining that two twins (i.e. two persons with exactly the same age) have different history. One of the twins leaves the earth with velocity near the light velocity, travels toward a star and comes back to earth. In the meantime the other twin remains on earth. When the traveler twin arrives on earth, it appears that he/she is younger that the twin on earth. This thought experience was proposed by Langevin in 1911, more that hundred years ago. His provocative intention was to show that the new relativistic kinematics can have very surprising results and that the new concepts of space and time are different from the usual ones.

I do not know if Langevin was aware of the success of the experiment he proposed. Today the normal place of the twin paradox problem is only in textbooks since solutions of the problem are well established either in the frame of Special Relativity or that of General Relativity. However from the time of its first appearance (1911) until now, articles (see for example H.Lichtenegger and L.Iorio (2011)) were and are regularly published on this subject. In general these articles intend to present new and original solutions or merely repeat what was already published. This continuity in publishing such articles is really puzzling. What is the reason to publish more and more papers on this theoretical subject? Rindler (2001) wrote: "Reams of literature were written on it unnecessarily for more than six decades. At its root apparently lay a deep psychological barrier to accepting time dilatation is real. From a modern point of view, it is difficult to understand the earlier fascination with this problem or even to recognize it as a problem." p.67). Following Grøn (2006) (who wrote himself a paper on the subject) more than three hundred articles were published on the twin paradox!

The situation is really complex and one can ask why people proposed so many solutions? Furthermore, why the solutions are so complicated, even in textbooks? One has the feeling that a too simple solution will destroy the mystery of the twin paradox.

This problem was also at the center of harsh controversies among physicists about Relativity and its place as a science. All this shows that this phenomenon (the paradox itself and all its manifestations) might have roots in the unconscious mind of human beings.

I thought that it was worth to investigate this strange situation and try to understand this fascination. The paper is divided in two parts. In the first, I describe with more details the history of the "paradox" and its solutions. It appears that there are some errors in many articles and this why I recall some results in Special Relativity. This will help in understanding the solution proposed in this paper that I shall call the simplest solution. First I analyze the Langevin article and then I define precisely the problem. In fact, it is possible to see that there are two versions of this experiment and the distinction is not done in the majority of the articles. I begin the second part in discuting some of the proposed solutions, mentioning the controversies and trying to understand the issue of these disputes. I present also some speculations which can help to decipher what are the possible psychological mechanisms behind the phenomenon: why this obsession? Why too many complicated solutions? Why the controversies? Clearly I am not able to prove the assertions relatively to the phenomenon but I hope that they will suggestive enough to help understanding it. One can already note that Langevin did not write about a paradox or twins as we shall see below.

The first mention of the twins in the context of the paradox is probably due to Weyl in his book *Space-Time-Matter* published in 1922. At the page 187 he writes: " Suppose we have two twin-brothers who take leave one another at a world-point A, and suppose one remains at home (that is, permanently at rest in an allowable reference-space), whilst the other sets out on voyages, during which he moves with velocities (relative to "home") that approximate to that of light. When the wanderer returns home in later years he will appear appreciably younger than the one who stayed at home." It seems that during a long time the name "twin paradox" was forgotten and people used only "clock paradox" as it is possible to see from the books and articles published before 1955: books of Tolman (1934) and of Møller (1952), article of Hill (1947) and that of Milne and Whitrow (1949). In all these texts, only appear the words "clock paradox".

Some authors relate the problem to Einstein in his famous paper of 1905 (Einstein (1905)). It concerns two clocks, one at rest and the second making a closed path. It is a slightly different problem and I shall discuss it later.

# Part one: two twins with two or three clocks

## I 1. Some results in Special Relativity

The relativistic kinematics concerns systems which move relatively one to another with constant velocity (in absolute value and in direction). For the time these systems never stop, move indefinitely and are called inertial systems. Later I shall discuss what happens if one system will stop. Inside a system, an event is defined as: "something that happens in a place defined by its Cartesian coordinates and by a time given by clocks". It is possible to imagine that in each place of the system there is a clock such it is possible in all places to know the time. All the clocks are at rest and it possible to synchronize them all together.

*1.1 Lorentz transformation*

Now consider two reference systems S and S' such that the axis xyz of S are parallel to those of S' x'y'z'. In S the time is measured as t and in S' as t'. The two system move relatively each to other with velocity V parallel to x and x'. S' moves relatively to S with velocity V and S move relatively to S' with velocity – V. If an event takes place in S at (x,y,z,t) what are the values of (x'y'z't') of this event measured by an observer in S'? The answer is given by the Lorentz transformation. Supposing that an event can happen only along the x and x' axis, the Lorentz transformation can be written as

$$x' = \gamma(x - Vt) \qquad (1)$$

$$t' = \gamma(t - Vx/c^2) \qquad (2)$$

where c is the light velocity, $\gamma = (1 - \beta^2)^{-0.5}$ and $\beta = V/c$. In (1) and (2) it is implied that in the two systems S and S', there is coincidence of their origin O and O' (such that x = x' = 0) when the clocks in S and S' indicate the same time taken equal to 0. It is also possible to take choice of x and x' different from zero, for t = t' and a constant must be added to the right side of (1).

Form (1) and (2), one can obtain the value of x and t corresponding to an event (x', t') taking place in S'

$$x = \gamma(x' + Vt') \qquad (3)$$

$$t = \gamma(t' + Vx'/c^2) \qquad (4)$$

*1.2 Applications of the Lorentz transformation*

Now one applies the Lorentz transformation to the following cases. First, one considers a process compound of two events in the frame S'. One event is the beginning of the process and takes place in S' at (x' = 0, t' = 0), and the a second (which is the end of the process) takes place at (x' = 0, t' = $\tau_1$) i.e. at the same place. The total time of this process is $\tau_1$ in S'. What is the time $\tau_2$ of the process measured in S? From (3) and (4), one gets for the end of the process, x = $\gamma V \tau_1$ and $\tau_2 = \gamma \tau_1$. In other terms, the time measured in S is larger than that measured in S' since $\gamma > 1$.

Conversely, consider now the same process which takes place in S between the events (x = 0, t = 0) and (x = 0, t = $\tau_1$) i.e. the time of the process is $\tau_1$. In S' using (1) and (2), one finds that the end of the process is measured in S' at (x' = $-\gamma V \tau_1$, $\tau_2 = \gamma \tau_1 > \tau_1$). One notes that the time measured in the system in which the process occurs is smaller than measured in another system moving relatively to the first. But the distance measured for the traveling of the origin is the same in the two cases, there is no "length contraction".

Consider again the two systems S and S'. Suppose that the same process takes place in both systems. At t = t' = 0 the two origins O and O' coincide and this is the first event in the systems : (x = 0, t = 0) and (x' = 0, t' =0). The identical process takes place at O and O' and its duration is $\tau$. The second event is in S at (x= 0, t = $\tau$) and in S' at (x' = 0, t' = $\tau$). From what it was concluded above, in S an observer measures $\tau$ for the process in his own system (S) and $\gamma \tau$ for the process in S'. Reciprocally in S' an observer measures time $\tau$ for its own process and $\gamma \tau$ for the process in S. This result is the basis for the clock or twin paradox. One can consider that $\tau$ is the elapsed time for the travel for each twin in his own system (the earth or the missile) and each of them measures the time $\gamma \tau$ for the other. The question is: what times are measured in both systems when they meet again?

A third case is interesting to investigate. The system S' moves away from S at the velocity V and one measures the position of O' in the two systems. The first event is define by x = x' = 0 and t = t' = 0. Second event in S is as follows: the origin O' performs a distance L during a time equal to L/V i.e. the event is defined by (x = L, t = L/V). What are the coordinates of this second event in S'? From (1) and (2), one gets (x' = 0, t' = L/($\gamma$V)). x' is zero since it is the position of O' in S' and the time in S' is smaller than in S.

Now one considers the same process but from the point of view of S' in which an observer measures that O goes away with velocity –V. In S', the position of O is given by the distance L' and the time is L'/V. In S, using (3) and (4), one gets x = 0, and t = L'/(γV). What is the relation between the two distances L and L'? Since there is a complete symmetry between the two points of view in S and in S', the times measured in S and S' must be equal. It results that L = L' and in particular there is no "length contraction".

*1.3 The proper time in Special Relativity*

Suppose that a particular process takes place in an inertial reference frame. The process is seen as a series of successive events. At each instant t (in this inertial frame) the process is characterized by an event located at x (I suppose that the process takes place only in the x direction). Between two very near events (t,x) and (t+dt, x+dx), the quantity $D^2 = [(dt)^2 - (dx/c)^2]^{0.5}$ is an invariant i.e. it takes the same value in all inertial references. One define the instantaneous proper time dτ as dτ = $[(dt)^2 - (dx/c)^2]^{0.5}$. It can be written also as dτ = $dt[1-(dx/cdt)^2]^{0.5}$. This particular time defines the proper time of an inertial reference frame moving with velocity (dx/dt) and with origin located at the point x. One sees that the process goes with a succession of inertial frames and in fact they can be seen as the successive positions of a unique non-inertial frame. If now one considers the whole process between two times $t_1$ and $t_2$ and between two positions $x_1$ and $x_2$, the proper time is given by

$$Dτ = \int_{t1}^{t2} \sqrt{1 - V(t)^2/c^2}\, dt \qquad (5)$$

with the integral being taken between $t_1$ and $t_2$. This proper time is independent of the frame where the measurement of $t_1$ and $t_2$ were made. In other words, in another frame different from the first one, the proper time is the same. (5) gives the possibility to calculate the proper time of some process in any frame one wants. In particular there is always a frame in which the process takes place at the same value of x and the proper time indicates the time of the process in this particular frame.

In the case where (dx/dt) is constant, the process takes place in the same place in an inertial reference frame which moves with the velocity V = (dx/dt). In such case one has merely Dτ = $Dt[(1-(V/c)^2]^{0.5}$ where Dt is the time elapsed in the process in the original inertial reference frame. Considering the heart of a human being as a clock, one sees that the proper time of this human being is his age. We shall see below how to use this concept of proper time.

*1.4 When an inertial frame stops*

One considers again two reference frame S and S' with a relative velocity V. Each of them has an identical clock which gives the time in each frame. In the frame S' there is a motor engine which is able to change its velocity relatively to S. Suppose now that people in the frame S' decide to activate their motor engine such that their velocity relatively to S becomes zero. In other words, the frames S and S' are now the same reference frame. The velocity of S' relatively to S is equal to V between 0 and $t_0$ and is the decreasing function f(t) between $t_0$ and $t_1$ until that $f(t_1) = 0$. To fix the things, suppose that the two clocks were synchronized such that t = 0 corresponds to t' = 0. The people in S' activate their motor at the time $t_0$ measured in S and their relative velocity in null at the time $t_1$ measured also in S. The question is: what are the indications of the two clocks when they meet in S?

In S the time given by the clock is merely $t_1$ and in S' the clock will indicate the proper time of the frame S'. It is given by (3) when the velocity is measured in S. One has

$$t'_1 = \int_0^{t1} \sqrt{1 - V(t)^2/c^2} \, dt = t_0 \sqrt{1 - (V^2/c^2)} + \int_{t0}^{t1} \sqrt{1 - (f(t)^2/c^2)} \, dt \qquad (6)$$

It is not difficult to show that the integral of the right side of (6) is smaller than $(t_1 - t_0)$ because $[f(t)/c]^2$ is a decreasing function smaller than 1. Consequently $t'_1 < t_1$ and at their meeting, the clock of S' will indicate a smaller time than the clock of S. The exact value of the difference depends on the form of the function f(t) i.e. the deceleration of the frame S'. If there a sudden stopping in a very short time, f(t) is equal to zero and $t'_1 = t_1/\gamma$. It is also possible to imagine different scenarios for the meeting of the two frames S and S' with different results for the indications of the two clocks.

If now S is the reference frame which changes its velocity relatively to S' such that S "enter" the frame S', one have $t'_1 > t_1$. One can imagine a third possibility. It exists a frame in which S and S' have equal but opposite velocity. If now the two frames S and S' stop relatively to this third frame, their two clocks will indicate the same time.

The expression (6) is enough to understand the twin paradox as it is will be shown below. I take this occasion to stress that I did not find in any textbooks a discussion of this problem of the stopping of a frame and this would be a good introduction to the twins paradox.

In 1972 Hafele and Keating perform an experiment in bringing a Cesium clock around the world in leaving an identical clock at the starting point. The two clocks are synchronized and one is launched for its circular trip. At the end of the experiment the two clocks are in the same frame as at the beginning and their indication are compared. It appears that the traveling clock is behind the clock which stays on the ground and quantitatively the experimental results are in good agreement with the theoretical previsions. The difference with (6) is that one needs to include a gravitational effect due to the fact that the two clocks are not at the same height relatively to the ground.

*1.5 Final remarks*

I ask the reader to note two things. First I (and shall) consider always related events and not isolated events. Only the differences between the spatial distance and the temporal distance between two events have meaning in Relativity theory. Secondly, I do not use words like "the clock goes slower" or " the clock goes faster". This is to emphasize that the clock by themselves do not go slower or faster but measurements made in other reference frames (not the rest system of each clock) indicate that they go slower or faster relatively to the indications of the clocks in their proper rest systems.

The following sentence (Frye and Brigham, 1957) concerning the traveler twin is meaningless in the context of Special Relativity: "…by relativistic principles, at that speed [near that of the light], his watch, his heart beats, in fact his whole metabolism slows down to a rate determinate by the formula $t_A = t_B(1-v^2/c^2)^{-1/2}$ where $t_A$ is the time experienced by the twin who stays at home, $t_B$ is the dawn-out time caused by the rapid motion of the traveling twin..." From this sentence one understands that the motion itself induces a change in the rate of the clock and the heart of the traveler. Furthermore the word "caused" clearly indicates that the rate changes are consequence of the motion.

This false interpretation of the Special Relativity is analogous to the Lorentz-Fiztgerald (Lorentz (1923), Fiztgerald (1889)) hypothesis concerning the length contraction of body in motion relatively to the ether. In this hypothesis the motion itself induces the change in the length of the body. As well known in the Special Relativity theory, they are the measurements of the length of a rod in a different reference frame system (not that of the rod itself in which one can define its proper length) that give a length different from the proper length.

## I 2. The Langevin article

The article of Langevin was written for a non-specialist audience which wants to know what it is the Relativity theory. It is only Special Relativity since the General Relativity was not known at this time. It does not contain any formulas or mathematical symbols. In a first part, Langevin presents the development of the Relativity theory from the Principle of relativity applied to electromagnetism until he shows the new conceptions of space and time. This very clear exposition of the Relativity is not completely in accordance with Einstein since Langevin proposes not to abandon the concept of ether (in fact he does not mention Einstein but only Lorentz). Langevin thinks that the presence of the ether could be proved through accelerated systems.

Now Langevin considers an accelerated system in which two events take place and concludes that the time between the two events in this system will be shorter than that measured by a "group of observers associated with a reference system in an arbitrary uniform motion". He goes on in taking the case "a portion of matter" making a close path and concludes: "And we can say that for observers related to that portion of matter, the time period elapsed between the departure and return i.e. the proper time of the portion of matter will be shorter than for observers who would have stayed connected to the reference system in uniform motion. This portion of matter will have aged less between its departure and its return than if it had not been accelerating". It is exactly the remark already made by Einstein in his 1905 paper.

And he notes with humor: "We can still say that it is sufficient to be agitated or to undergo accelerations, to age more slowly ". Langevin presents the preceding result in a different light: "Suppose that two pieces of matter meet for the first time, separate and meet again. We can say that observers attached to the portions during separation, will not evaluate that duration in the same way, as some have not aged as much as the others". Always with humor he goes on: "This remark provides the means for any among us who wants to devote two years of his life, to find out the Earth will be in two hundred years, and to explore the future on Earth, by making in his life a jump ahead that will last two centuries for Earth and for him it will last two years … For this it is sufficient that our traveler consents to be locked in a projectile that would be launched from Earth with velocity sufficiently close to that of light but lower, which is physically possible, while arranging an encounter with, for example, a star that happens after one year of the traveler's life, and which sends back to Earth with the same velocity. Returned to Earth, he has aged two years, then he leaves the ark and finds our world two hundred years older…"

This is the first mention of the thought experience that one cannot call "twin paradox" since for Langevin there is no paradox nor twins!

Langevin imagines that the traveler and people on earth can send electromagnetic signals and this also the origin of the solutions using light signals as exposed in several books (for example Bohm (1965), Resnick and Halliday (1972))

I quote again Langevin: "Thus the asymmetry – which occurred because only the traveler, in the middle of his journey, has undergone an acceleration that changes the direction of his velocity, and which brings him back to his starting point on Earth – results in the fact that the traveler sees the Earth moving away and approaching during times that equal to one year, while the Earth, which is informed of the acceleration only by the arrival of light waves, sees the traveler moving away during two centuries and returning during two days, i.e. during a time forty thousand times shorter." The end of the article is devoted to practical realizations of the experience, always with humor.

It is clear that Langevin is aware of the reciprocal measures of the travelling time and in fact choosing the ratio of the two times to be equal to 100 means that for him $\gamma = 100$ or $\beta = 0.99995$. However he attributes the time asymmetry to the acceleration. It is also the beginning of a controversy if the acceleration is essential in order to explain this thought experiment.

For a better understanding of Langevin, it is useful to consider the paths of the Earth and of the traveler in spacetime diagram (Fig.1). The line OA is the path of observer who stays in the Earth, the line OB is the path away from the Earth for the traveler and BA is his return path. Because the time dilatation OA corresponds to 200 years but the lines OB and BA to one year respectively. The dash lines BK and BL are the light paths of signals sent by the traveler. The signal BK is the last signal sent just before his return move and BL is the first signal he sends just after his return move. Following Langevin, OK and LA correspond at two days (of course the figure cannot be at scale). The asymmetry is well represented in this diagram and in particular one sees the jump in the time measured by the traveler for observer on Earth during his return move.

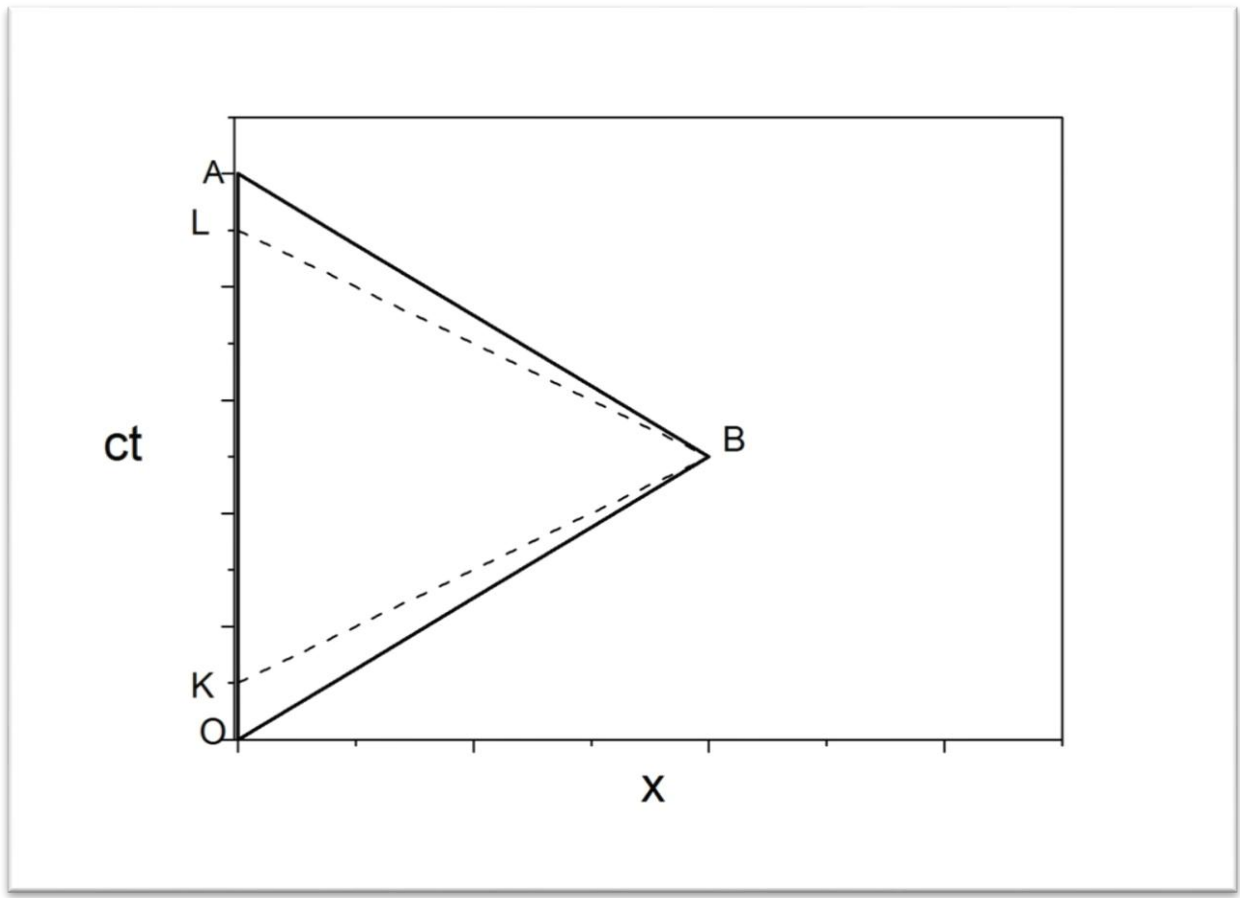

Fig.1 Spacetime diagram of the Langevin traveler. OA is the trajectory of the earth and the lines OB and BA that of the traveler. BK is the last light signal received by the earth when the traveler reaches the star and BL is the light signal sent by the traveler when he turns out at the star. For the earth inhabitants the return time of the traveler is only the distance LA. The time OB is one year like the distance BA but the distance OA is 200 years.

I 3. The problem

In the literature on the twin paradox, one finds a controversy if acceleration an essential ingredient of the problem i.e. to find the good solution, that given by Langevin: the traveler is younger that the stay at home when he comes back. One has a very special situation. It seems that the majority of the scientists have accepted the point of view of Langevin (but not all, see below). In spite of the variety of solutions presented during the last hundred years, all solutions are different from that of Langevin! Thus all people presenting a solution know in advance the correct answer. It is exceptional in Physics that one tries to solve a problem when the solution is already known.

If one wants to imagine the journey of the traveler, the things can be seen as follows. In a first part, the traveler, at the beginning at rest, is accelerated until he reaches the velocity V. In a second part, the motor of his engine is off and he continues his way with a constant velocity V. In a third part, a force is applied in order to decrease his velocity until it becomes null. In the three following parts of the return, one has increase of the velocity, constant velocity and finally decrease of the velocity until it becomes zero. At this point, he puts off the motor of his missile and meets the Earth observer. Both are now at rest in the same frame and are able to compare the readings of their clocks. I shall call this version of the journey, the true story. The story is characterized by the fact that the traveler, at least in some parts of his trip, is accelerated in activating his engine.

Since the reference frame of the traveler is accelerated relatively to the reference frame of the earth, it is not an inertial reference frame. Nevertheless, the resolution of the problem can be made with the help of the Special Relativity in the frame of the stay at home but also a resolution with the help of the General Relativity theory in the frame of the traveler will give more insights as one can see that below.

Now, it is possible to invent a second scenario also compatible with the Special Relativity but without intervention of neither accelerations nor decelerations. The traveler gains his velocity in the region $x < 0$ and moves in the direction $x > 0$ with constant velocity V. He passes in front of the earth observer located at $x = 0$ with that velocity. At the coincidence of the two observers, they adjust their clocks to zero. The traveler goes on until he crosses another missile moving with velocity $-V$ exactly at the point he wanted to reach. Now, he jumps into this new missile to come back to earth and the clock of the reference frame of this new missile is adjusted to that of the first missile. When he passes again in front of the earth observer, they compare their clocks. The traveler does not begin his motion at $x = 0$ nor stops at $x = 0$ when he comes back. In this version of the problem, no acceleration appears. In this scenario there are two twins, but one notes that there are three reference frames and three clocks! This scenario of the three reference frames was already proposed by several authors. See the Arzelies's book (1966) and references inside.

What is the criterion which permits to decide if the traveler is younger when he comes back? The problem must be divided in two parts. In the first, one considers the point of view of the observer who stays at home on the earth: he is at rest in his own frame. He measures the travel time in his frame and determines the travel time that the traveler will measure. It is possible to calculate the proper time of the traveler or to use the Lorentz transformation. In a second part, one considers the point of view of the traveler which is at rest in his own frame. Now from his point of view the earth and the star are moving: first the earth goes away and the star comes near the traveler and in the second part, the earth becomes nearer and the star goes away. The traveler measures the travel time of the earth or the star and determines the time measured by the stay at home which is for him in motion. If a) the stay at home finds that his time is larger than that for the traveler b) the traveler finds that the time in his frame is smaller than that in the frame of the stay at home, both agree that the journey time for the traveler is smaller than for the stay at home. The reciprocity evoked above is lost.

To be more specific it is useful to determine four times. In the frame $S_1$ of the stay at home, the observer measures his own value $(\Delta T)_1$ of the round trip of the traveler and he determines what is the time $(\Delta \tau)_1$ measured by the traveler corresponding to $(\Delta T)_1$. In the frame $S_2$ of the traveler, the traveler measures his own value of the round trip $(\Delta \tau)_2$ and he determines what is the time $(\Delta T)_2$ of the earth observer corresponding to $(\Delta \tau)_2$.

## I 4. The solutions

*1. The solution of the true story: Special and General Relativity*

The complete solution was given by Møller (1952). This solution was also presented by (among others) Leffert and Donahue (1958), Perrin (1979) Styer (2007), Müller, King and Adis (2008), Already in 1934, Tolman has already given a simplified solution and in the light of the work of Møller, it is appears that it is only a partial solution. Several other articles present also the solution of Tolman (Frye and Brigham (1957), Cochran (1960), Grøn (2006))

I shall present first the solution of Møller which is well detailed. It includes two steps: one based on the Special Relativity and the second the General Relativity. The journey of the traveler toward the star is compound of three parts (fig.2). In the first part, between the points E (the earth) and B, a constant force F is applied to the traveler until he reaches the velocity V at the point B. In the second part he goes on with constant velocity V between the points B and C. Finally at C the same force F is again applied but in the direction opposite to his motion (i.e. x < 0) in order to decrease his velocity. At the point S (the star) the velocity is null and his motion is reversed, always under influence of the force –F.

The return travel is symmetric to the first path. Along SC the force –F is applied in the direction of the motion which is now accelerated. At C, he reaches again the velocity V (and by symmetry one has EB = CS) and finally at B the force F is applied in the direction x > 0 in order to decrease the velocity to zero at the point A. It is clear that the motion is the four portions EB, CS, SC and BE is identical.

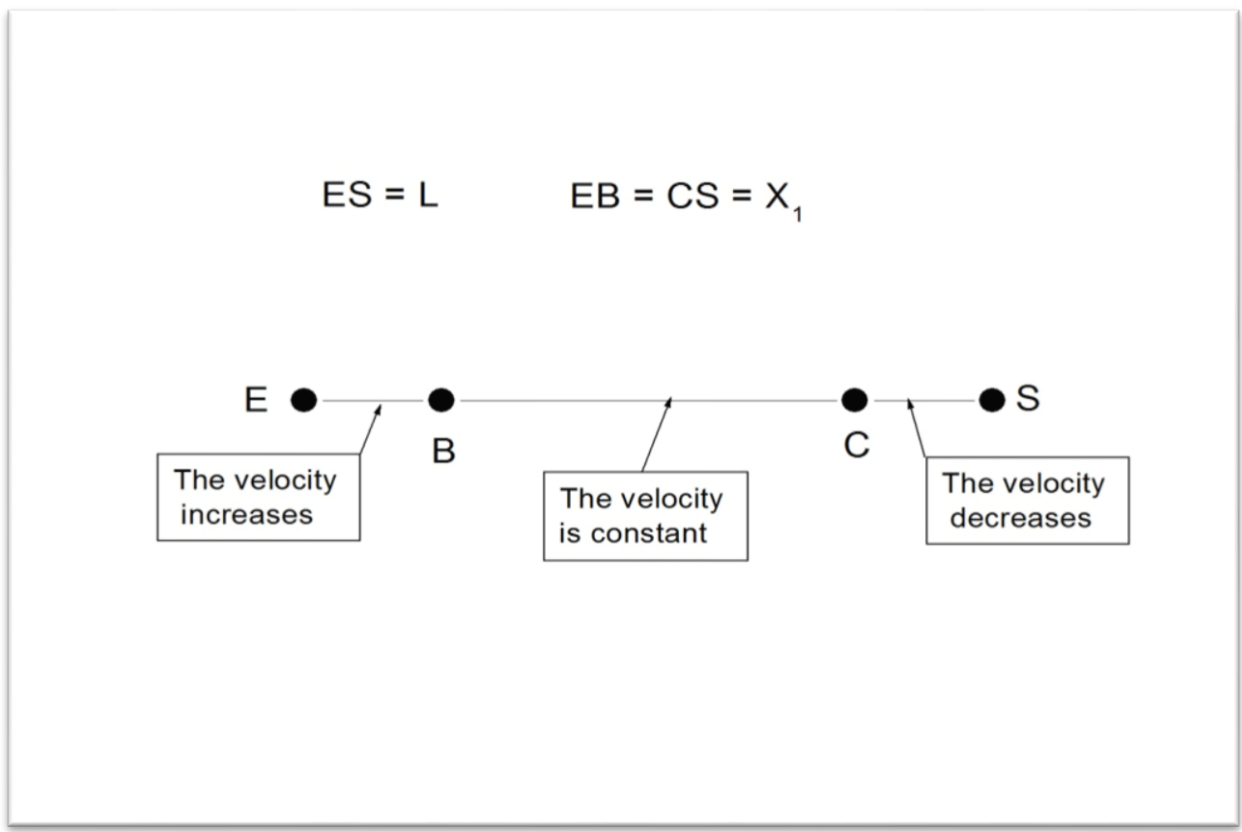

Fig.2 Travel from the earth E to the star S in three parts: EB increasing velocity under the application of the force F, BC constant velocity (F = 0) and CS with the force –F and decreasing velocity until it becomes null. The back travel corresponds to: SC increasing velocity, CB constant velocity and BE decreasing velocity

One begins by calculate the travel times in the frame of the earth observer $S_1$. The motion of the traveler between E and B is easy to calculate. One applies the relativistic dynamics

(Møller (1952))   to a mass M (mass of the traveler) under a constant force F.   The equation of the motion is

$$d[u/\sqrt{(1-u^2/c^2)}]/dT = F/M \qquad (6)$$

where u = dX/dT and  X is the displacement. A first integration gives the velocity and one gets, assuming that the velocity is null at t = 0

$$u = gT/\sqrt{[1 + (gT/c)^2]} \qquad (7)$$

g is a parameter[1] equal to F/M. A second integration gives the displacement X(T) as a function of T

$$X(T) = (c^2/g)\{ [1 + (gT/c)^2]^{0.5} - 1 \} \qquad (8)$$

and for the velocity V of the traveler at the time $\Delta'T$ when he reaches the point B

$$V = g\Delta'T/ [1 + (g\Delta'T/c^2)]^{0.5} \qquad (9)$$

One deduces form (9)

$$g \Delta'T = V/[1 + (V/c)^2]^{0.5} \qquad (10)$$

One can rewrite $\Delta'T$ as equal to $(c/g)\beta\gamma$ and the distance EB as $X_1 = (c^2/g)(\gamma - 1)$. The time between B and C is

$$\Delta''T = (L - 2X_1)/V = L/V - (2c^2/gV)(\gamma - 1) \qquad (11)$$

where L is the distance ES between the earth and the star.  The elapsed time for the travel E to S is

$$(\Delta T)_1 = 2 \Delta'T +  \Delta''T = L/V + (2c/g)(\gamma - 1)/\beta \qquad (12)$$

The travel time of the distance ES in the frame $S_2$ of the traveler is his proper time

$$(\Delta\tau)_1 = \int_0^{\Delta T} \sqrt{(1 - (\tfrac{V}{c})^2)}\ dt \qquad (13)$$

From (11) and (12) giving $\Delta'T$  and  $\Delta''T$ one gets from  (13)

$$(\Delta\tau)_1 = [L/V - (2c^2/gV)(\gamma - 1)]/\gamma + (2c/g)\text{th}^{-1}\beta \qquad (14)$$

---

[1]  This parameter g is not the acceleration but it has only the dimension of an  acceleration.

Now one supposes the frame $S_2$ at rest and one calculates the travel time is this frame and that of the stay at home observer. This method was already proposed by Einstein in 1918 (Pesic (2003)). In the time intervals where $S_2$ is accelerated relatively to $S_1$, there is a gravitational field in $S_2$. The gravitational potential is given by $\chi = -gx(1 - gx^2/2c^2)$. From the theory of the General Relativity, two clocks separated by a distance x in the direction of a gravitational field have different rate such that the times are related by $t_2 = t_1[1 + \chi(x)/c^2]$ where the clock at the lower potential having the smaller rate (Tolman (1934), Møller (1952), Cochran (1960)). This the physical basis for the determination of the travel time in the frame $S_2$.

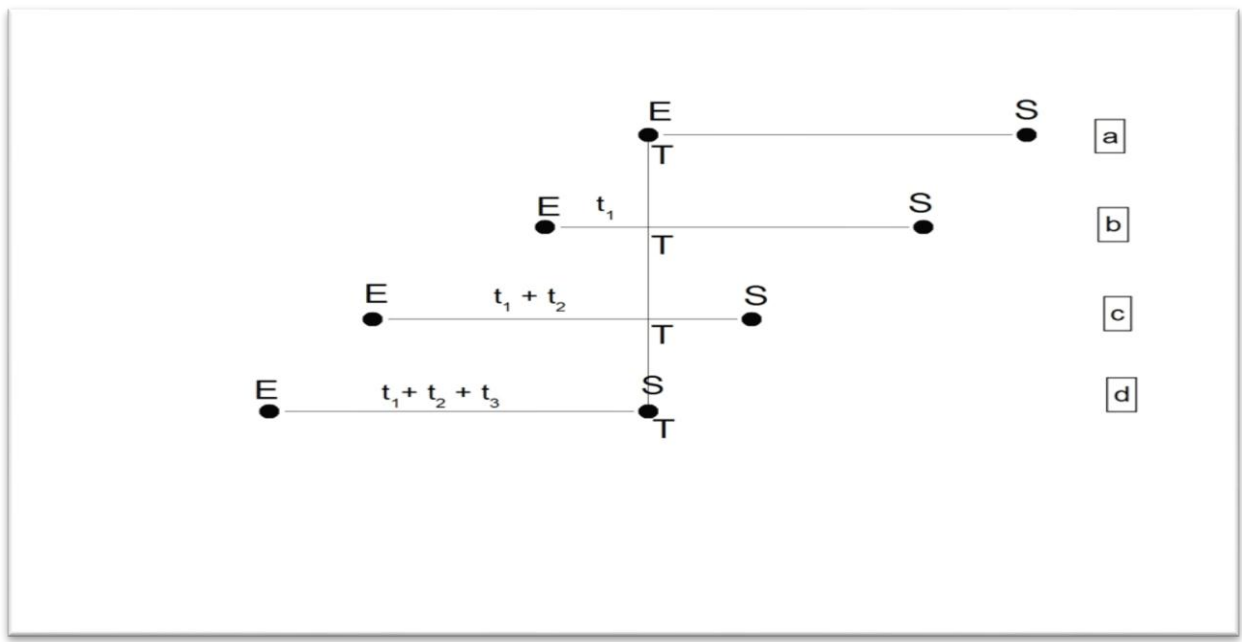

Fig.3 Positions of the earth and the star from the point of view of the traveler T. a) At the time t = 0  b) At the end of the time where the traveler (in the earth frame) is accelerated c) At the end of the time where the traveler has moved with constant velocity (in the frame of the earth) d) The star has reached the traveler.

In fig.3, are indicated the different steps of the first part of journey as seen from the traveler which is supposed to be at rest in his frame. Fig.3a corresponds to the starting time when the earth E and the star S after time $t_1$ where in the frame of the earth the traveler is accelerated. In the frame $S_2$ the earth is falling freely in the direction of the negative x. Fig.3c describes the state after the time $t_1 + t_2$ where $t_2$ is the time where the earth moves with the constant velocity V. Finally, the star S reaches the traveler and the earth is at the largest distance. The time elapsed from the beginning is $t_1 + t_2 + t_3$ where $t_3$ is the time where again there is a gravitational field.

The travel time measured by the traveler in his own frame $S_2$ is given by

$$(\Delta\tau)_2 = [L/V - (2c^2/gV)(\gamma - 1)]/\gamma + (2c/g)\text{th}^{-1}\beta \qquad (15)$$

identical with $(\Delta\tau)_1$. The elapsed time $(\Delta T)_2$ at earth measured in the frame of the traveler is given by

$$(\Delta T)_2 = L/V + (2c/g)(\gamma - 1)/\beta \qquad (16)$$

which is identical with $(\Delta T)_1$ the time elapsed on earth measured by the stay at home observer. In other words, the times measured in both frames $S_1$ and $S_2$ are the same. This result was expected. Since the traveler and the stay at home are in the same reference frame at the beginning and at the end, one has to find the same age for each whatever the method of determination.

First, one considers the limit $g \rightarrow \infty$, and gets that the ratio of the times $(\Delta T)/(\Delta\tau)$ is equal to $\gamma$ i.e. the total time measured by traveler is smaller than the time measured by the stay at home traveler.

To have a better understanding of this last result one considers the time in $S_1$ as measured by the traveler in his frame $S_2$. It is compound of three parts (for the sake of simplicity only the times in the path outbound are presented). The first correspond to the acceleration period in $S_1$ or with presence of the gravitational field in $S_2$. It is equal to $T_1 = V/g$, the second in the region where the velocity is constant, is equal to $T_2 = \Delta''T/\gamma^2$ with $\Delta''T$ is given by (8) and in the region of deceleration, it is equal to $T_3 = (c/g + L/c)(V/c)$. If $g \rightarrow \infty$, the distances EB and CS (fig.2) shrinks to zero and the time $T_1$ also goes to zero. However, the time $T_3$ does not go to zero but to $(LV/c^2)$. In other words, the time at the turning point, in spite of the fact it takes place in a zero distance, is not zero when measured in the stay at home frame! It results that the total time $\Delta T$ exhibits a discontinuity at the turning point. This is the reason of the dissymmetry in the times of the two twins. Møller explains this particularly surprising result by the fact that in such a case ($g \rightarrow \infty$) the gravitational potential is infinite. This result is very important because it shows that, even if the acceleration times are very short, the asymmetric aging exists. It is very often claimed that, it the acceleration times are very short, the acceleration becomes irrelevant to the problem and this is false.

It is also interesting to consider the case where g is different from infinity. The ratio K of the travel times for the stay at home and the traveler is given by $\Delta\tau/\Delta T$ and one asks if this ratio which goes to $\gamma^{-1} < 1$ for $g \to \infty$, may be larger than 1 or is always smaller than 1? If it is larger than 1, it should mean that the stay at home may be younger than the traveler for some values of g. The answer is given by an inequality

$$(Lg/2c^2) > (\beta \, th^{-1}\beta)/(1 - 1/\gamma) - 2 = Q(\beta) \qquad (17)$$

If the inequality (17) is satisfied, then K < 1 and after his trip, the traveler arrives younger than the stay at home. However a second inequality must be verified. The acceleration distance $EB = X_1$ and the deceleration distance $CS = X_1$ cannot be larger than L/2. This gives a second inequality

$$(Lg/2c^2) > (\gamma - 1) = P(\beta) \qquad (18)$$

In fig.4 are drawn the functions $P(\beta)$ and $Q(\beta)$. One sees that the curve $P(\beta)$ is always above the curve $Q(\beta)$. This means that if the travel is possible i.e. $2X_1 \leq L$, the inequality (14) is always satisfied and K < 1. The traveler is always younger than the stay at home whatever the value of g but the ratio K may be different from $1/\gamma$.

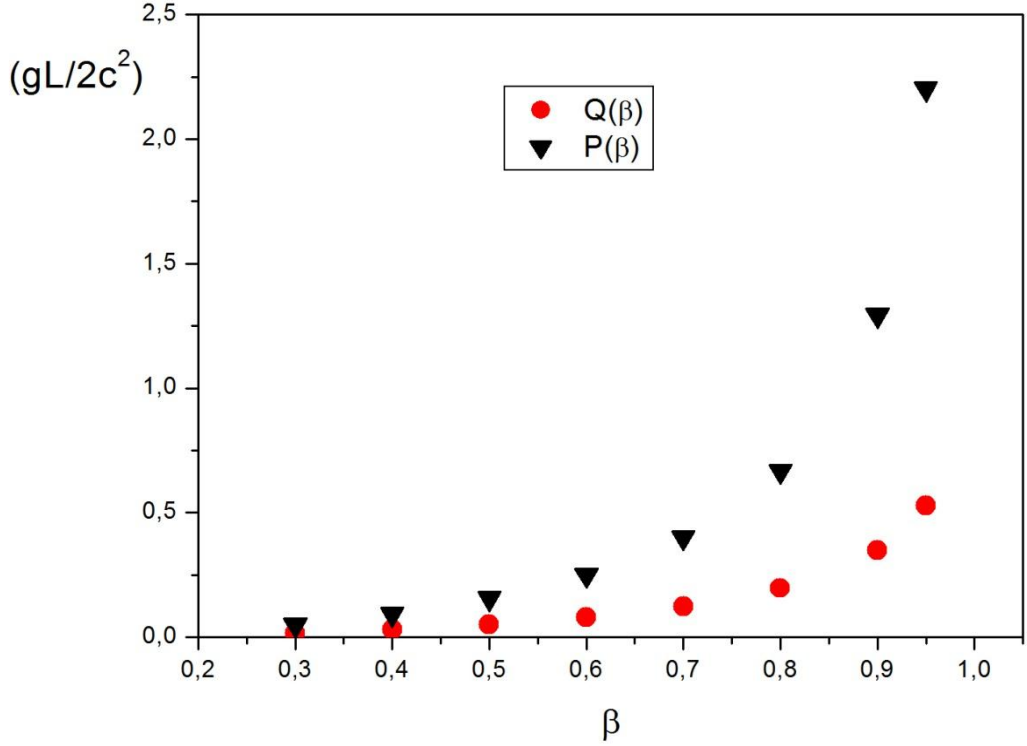

Fig.4 (see text)

Finally we add some words about the Tolman's solution. It is supposed that the acceleration distances and deceleration distances are very small (this is equivalent to take g infinite). In the frame $S_1$, one uses only the Special Relativity and one gets the same results as above i.e. $(\Delta T)_1 = L/V$ and $(\Delta \tau)_1 = L/V\gamma$. In the frame $S_2$, one introduces the influence of the gravitational potential of two clocks such that the relation of the two clock rates in a gravitational field is $t_2 = t_1(1 + \chi/c^2)$. As mentioned above, when in the frame $S_2$, one determines the variation of the time of the earth observer during the journey, one gets a discontinuity. During the first part the time increased from zero to $(L/V\gamma)$, during the very rapid change in its direction, the time increases suddenly by $(LV/c^2)$ and in the return trip it increases again by $(L/V\gamma)$. Taking $1/\gamma = (1 - \beta^2)^{0.5}$, one has for the total travel time $2\Delta T$ for the earth observer (as determined from the frame $S_2$)

$$2(\Delta T)_2 = 2L/V\gamma + LV/c^2 \qquad (19)$$

Only if β is small enough, $1/\gamma \approx (1 - \beta^2/2)$, one gets for $(\Delta T)_2$

$$2(\Delta T)_2 \approx 2(L/V)(1 - \beta^2/2) + (L/V)(V/c)^2 = 2\, L/V \qquad (20)$$

The ratio $K = \Delta\tau/\Delta T$ is equal to $1/\gamma$ only for small velocity (taking $\Delta\tau = L/V\gamma$). In view of the results of Møller, one concludes that the Tolman solution is only approximate and good for small velocities.

To close this section I present two remarks. First the calculation of the travel times in the frame $S_1$ of the stay at home observer were made completely in the frame of the Special Relativity and only in the calculation in the traveler frame $S_2$, one uses the General Relativity. Secondly, the Møller solution corresponds to a special case where the force F is applied suddenly and remains constant. Other cases are possible and the solutions may be very different (Wu and Lee (1972)).

## 2 *The solution without acceleration*

The stay at home is in the A frame and the traveler begins his trip associated with the frame B which passes in front of A with the velocity V in the $x > 0$ direction. After the time $\tau_0 = L/V$ he crosses the new frame C which moves in the direction of the earth with velocity $-V$.

One begins by determination of the times from the point of view of the stay at home. At the beginning the position of the traveler is given by: ($x_A = 0$  $t_A = 0$) in the frame A and ($x_B = 0$  $t_B = 0$) in the frame B. After the traveler (frame B) has passes the distance L one has for the position of the traveler, in the frame A ($x_A = L$  $t_A = \tau_0 = L/V$) and in the frame B

$$x_B = \gamma(x_A - V t_A) = 0 \quad \text{and} \quad t_B = \gamma(t_A - V x_A/c^2) = \gamma\tau_0(1 - V^2/c^2) = \tau_0/\gamma \qquad (21)$$

For the trip of the frame C which comes back to A, one gets the same results: for A the time is also $\tau_0$ and for the frame C it is also $\tau_0/\gamma$. The total time of the travel is: for the stay at home is $2\tau_0$ and for the traveler $2\tau_0/\gamma$. One sees that for the stay at home the travel time is larger than that of the traveler.

Consider now the frame B (it is now at rest) and one determines the time for A to make the distance $-L$. At the beginning one has again ($x_B = 0$  $t_B = 0$) and ($x_A = 0$  $t_A = 0$). When A reaches the distance $-L$, one has ($x_B = -L$  $t_B = \tau_0$) by reciprocity. Now for A one has

$$x_A = \gamma(x_B + V t_B) = 0 \quad \text{and} \quad t_A = \gamma(t_B + V x_B/c^2) = \gamma\tau_0(1 - V^2/c^2) = \tau_0/\gamma \qquad (22)$$

as expected by symmetry.

Now one considers the frame C when the frame A moves in direction to C. When the frame A begins his motion toward the frame C, one has ($x_C = -L$   $t_C = \tau_0$). But for A one has

$$x_A = \gamma (x_C - V t_C) + K = 0 \qquad t_A = \gamma (t_C - V x_C/c^2) = \gamma \tau_0 (1 + \beta^2) \qquad (23)$$

One must add a constant K in order to get $x_A = 0$. At the arrival of A at C, one has for C ($x_C = 0$   $t_C = 2\tau_0$), and for A one has

$$x_A = \gamma (x_C - V t_C) + K = 0 \quad \text{and} \quad t_A = \gamma (t_C - V x_C/c^2) = 2\gamma \tau_0 \qquad (24).$$

One verifies that the constant K is equal to $2V\gamma\tau_0$ in (20) and (21). For A the total travel takes the time $2\gamma\tau_0$ and for the traveler $2\tau_0$. Here again the time of the traveler is shorter than that of the stay at home. It is also possible to calculate the time for the frame A (in the point of view of the frames B and C) to come back to its original position by subtracting (20) from (21). One gets $\tau_0/\gamma$. Thus for the stay at home, the time from the point of view of B and C is divided in three parts: first when A goes away, time $\tau_0/\gamma$ from (19), the change of frame from B to C induces a change in the time $[\gamma\tau_0 (1 + \beta^2) - \tau_0/\gamma]$, and finally coming back of A to C with the time $\tau_0/\gamma$. One recovers the situation observed above, namely a jump in the time of A, always from the point of view of B and C. In conclusion, the reciprocity is broken. The final results are shown in the table 1

The preceding results namely that each twin has the same age when measured in his own reference frame was already obtained by A.F.Kracklauer and P.T Kracklauer (2000) and by E.Fischer (2010). These authors conclude that there is no twin paradox in the frame of the Special Relativity. The conclusion of A.F.Kracklauer and P.T Kracklauer is: "*Thus, when the whole trip is completed, both twins agree that they have experienced equal portions of proper time since the start of the trip. Their* reports *to each other via light signals on the passage of time, in the usual way do not agree, however.*" They are right but miss the points that the twins agree via their signals that the stay at home is seen older and the traveler younger.

|  | From the reference frame A | From the reference frames B/C |
|---|---|---|
| Travel time for the stay at home | $(\Delta T)_1 = 2L/V$ | $(\Delta T)_2 = 2\gamma L/V$ |
| Travel time for the traveler | $(\Delta \tau)_1 = 2L/\gamma V$ | $(\Delta \tau)_2 = 2L/V$ |

Table 1  The ratio $(\Delta T)/(\Delta \tau)$ = (travel time stay at home)/(travel time traveler)  is always equal to $\gamma > 1$.

## *3. Discussion*

The solution of the non-accelerated scenario presented above is not identical with that proposed in the literature.  It is almost always admitted that the traveler sees the distance Earth-Star shorter than that sawn by the stay at home by the factor $1/\gamma$. In other words, the reciprocity mentioned about results of the Special Relativity (see section 1.2) is overlooked.  I have already emphasized that in the first part of the trip (until the change of direction of the traveler) there is a complete reciprocity between the stay at home and the traveler. The consequence of this choice is that now the table 1 is transformed in the table 2.

One remarks that the twins agree completely on their ages. Their proper times (respectively $(\Delta T)_1 = 2L/V$   and $(\Delta \tau)_2 = 2L/\gamma V$)  are such that the stay at home is older even if they are not in the same reference frame.

|  | From the reference frame A | From the reference frames B/C |
|---|---|---|
| Travel time for the stay at home | $(\Delta T)_1 = 2L/V$ | $(\Delta T)_2 = 2L/V$ |
| Travel time for the traveler | $(\Delta \tau)_1 = 2L/\gamma V$ | $(\Delta \tau)_2 = 2L/\gamma V$ |

Table 2  The different times of the twins when the length contraction is introduced.

Now it is possible to understand the claim that the non-accelerated version can give the solution of the problem and the opinion that the acceleration is not a necessary ingredient for the resolution of the twin paradox. However, it is clear that the two situations are different. In the non-accelerated version, the twins are never in the same reference frame and thus there is no trip of one of them and no come back to Earth. It is not because the solutions of the two problems are identical that it is the same situation.

I found only one author (Grünbaum, (1954)) who discusses the equivalence of the non-accelerated solution  versus the true story  solution.  He admits that the solution of the three clocks (or the three frames) is different from that for a round trip of one clock, but nevertheless his conclusion is that they are equivalent.   Interesting enough, Grünbaum quotes one scientist who disagrees with him.  For Dugas (1954) "the clock paradox is a pseudo-problem from the standpoint of the special theory of relativity".  Dugas anticipated the conclusion of A.F.Kracklauer and P.T Kracklauer and that of Fischer: for Dugas it is a pseudo-problem, for A.F.Kracklauer and P.T Kracklauer it is a non-paradox and for Fischer it is a myth.

# PART TWO: Why?

It is now time to come back to the phenomenon itself and try an analysis.

## II 1 Solutions exist

In the first part I showed that effectively there are solutions of the problem of the twins paradox. If one wants to consider the problem in all its details (departure from rest, travel, change of direction, back travel and return at rest) there is the Møller solution (1952) or solutions based on the same principles but with a slightly different formalism (Perrin (1979)). The calculations made in the two frames, that of the stay at home and that of the traveler give identical results. In particular, the calculation made in the reference frame of the stay at home must be proposed to all students of the theory of the Relativity. This solution is in fact a special case (since the force is constant in the time intervals where it is applied) but Wu and Lee (1972) showed that the traveler is always younger whatever the acceleration applied to him. For those who prefer the scenario of non-accelerated frames, the solution presented above which is based on the Lorentz transformation is may be the simplest one. Thus the conclusion is clear: there are simple and complete solutions of the problem and they are easily accessible. In principle we should conclude that there is no place for further solutions and discussions.

## II 2 Disorder and frustration

Thus the surprise is great when one discovers the number of articles devoted to the subject. When one considers, at large, the ensemble of the solutions, the first impressions one has is "disorder" and "frustration". There are so many solutions such that one does know what to do. It is not clear if they are equivalent or may be some are correct and others not. Very often they are not simple but complicated, often very long and are all with the sole goal: to go to the known solution namely that the stay at home is older by the factor $\gamma$. The repetition of the same solutions in several articles is also misleading. And the controversy about the acceleration is also disturbing and not only for beginners.

The amount of papers is so large that it is not surprising to find among some authors unawareness about important articles. The most astonishing are those who do not know that the father of this thought experience is Langevin in 1911 but attribute it to … Gamow in a book of 1947! Others know the work of Møller only through his first publication in 1943 in a Danish journal and do note quote his book published in 1952. I have to confess that it is also possible that I missed some relevant works.

First it is astonished and strange that the first part of the solution of Møller which is entirely based on the Special Relativity is rarely found in the textbooks but more frequently in journals. Sometimes Einstein is quoted for his solution using General Relativity (the second part of the Møller work) but it is rejected in favor of the Special Relativity ignoring that a complete solution can be found in the frame of the Special Relativity (the first part of the Møller work). In the following I shall consider only the solutions which are claimed to be in the frame of the Special Relativity but different from that of Møller.

A typical example of confusion is given by Arzeliès (1966). First he uses the above result (eq.(6)) concerning the traveler's clock after stopping his motion and he does not give any explanation. To understand the point of view of Arzeliès one needs to add that the traveler has a null velocity twice: when he changes direction and when he stops on earth. As shown above when a frame stops, the clocks of the two frames do not coincide. For him this is the solution of the problem without trying to show where is a symmetry breaking. Secondly he thinks that the acceleration is essential even if it is applied during a very short time (as many assumed) and he does not give any explanation why. And thirdly, he does not agree that, from the point of view of the traveler, the elapsed time of the stay at home exhibits a discontinuity. These are too many reasons for creating trouble in the mind of the reader.

These are the following basis for the solutions and I give only some references:

The Lorentz transformation (Rossen (1964), Müller (1972))

The Minkovski diagram (Rossen (1964)) or the spacetime diagram (Chang (1993))

The Doppler effect and exchange of light signals (Bohm (1965) French (1968)

The three brothers or the three clocks (Grûnbaum (1954))

I shall not discuss all these solutions but I shall give a general overview. Some authors give the results only from the point of view of the stay at home and evidently one gets the desired results. How the symmetry is broken is not explained or sometimes just mentioned in saying that the traveler comes back in a different frame than that in the first part of his trip. Or in saying that the traveler experiments a force and for this reason the twins are not in equivalent positions. Of course, this is exact but there is no quantitative explanation which can help understanding the problem. There are relatively few papers or textbooks which give a clear and elegant exposition of the symmetry breaking. In this context I have to mention the paper of Chang in which the symmetry breaking in very well explained in taking the two points of view of the twins.

The place devoted to the twin paradox in several textbooks is relatively large, between 5 and 20 pages and the text is full of considerations on the controversy and how to resolve it. This long exposition is really an obstacle to an easy understanding of the problem. Furthermore some authors think that presenting several solutions may be giving a better insight but in fact they increase the confusion.

The goal is to find the ratio of the twins times equal to $\gamma$. Thus the final result is already known. For that it is necessary to accept the length contraction in the frame of the traveler, which is not exact. Finally it is necessary to mention that several solutions are too complicated. I think in particular on the solutions based on the exchange of signals between the twins. I wonder if it is really the simplest view of this problem. For example, French presents few explanations and a page full of expressions which, in principle, will permit the understanding of the problem.

Other examples of lack of clarity are the papers of Romer (1958) and Grünbaum (1954) who introduced several clocks or observers. Romer presents a long discussion without calculation in the hope that the reader will easily understand the origin of the dissymmetry. For that the author introduces not only the twins but also two other observers! Grünbaum introduces not only the three clocks of the model but also two other clocks!

The discussion on the role of the acceleration is, in fact, a vain controversy. The question is if the introduction of the acceleration is necessary to find the solution. In fact, since there are two scenarios for the paradox, there is no problem. In the first scenario the traveler experiences a force during some parts of his trip, and the acceleration is an integral part of the problem. In the second scenario, the acceleration is completely absent. I have already mentioned that the fact that the final results are analogous (but not identical) in the two scenarios, does not imply that they correspond to the same problem. It is why those who claim that acceleration is not necessary to solve THE problem merely ignore that there are two different scenarios. There is not ONE problem! A typical example of this confusion is given by Darwin (1957) in which he speaks about the reunion of the two ships but proposes a solution where there is no encounter of the ships! (by the way, his solution is not correct since there is not symmetry breaking in spite what he claims).

I have to add that the so frequent argument of a short acceleration times make them irrelevant is false. People think that it makes the problem easier but it is misleading. In some cases, the true story is eliminated and nobody knows how the traveler comes back.

Langevin attributes the difference in the twins' ages to the acceleration. It is possible that Langevin wanted to explain by the acceleration the symmetry breaking. But the complete calculations show the age difference is due to two factors: the acceleration and the time dilatation (see expressions (12) and (14)). I hope that the reader is convinced that the discussion about the role of the acceleration is unnecessary and the confusion about the two scenarios is removed.

I cannot close this analyze of the different approaches without mentioning Reichenbach. In order to explain the asymmetry between the twins, he proposes to introduce all the masses of the universe which are seen as fixed relatively to the stay at home. From his point of view, only the traveler is moving but from the point of view of the traveler the stay at home and the universe are moving making the two clocks inequivalent. Clearly it is not necessary to go so far in order to interpret the difference in the twins' ages.

To conclude this section, I express again my perplexity facing this chaos concerning the problem of the traveler with a velocity near that of the light. It is well known that there were and there are always opponents to the theory of Relativity. It is a pity to see that some circles take the traveler problem as evidence that the theory is false. I do not speak about physicists (see below about physicists) but about scientists in other fields who express their doubts about the validity of the theory of the Relativity. I feel that if Relativity physicists were presenting only simplest solutions, they would provide a very good service for the understanding of the Relativity.

## II 3 The deniers

It seems necessary to recall briefly the opinions of physicists who denied in some ways the results of the theory of the Relativity about the twin paradox. They are very serious physicists with an honorable record but they casted some doubts about the solution of the twin paradox problem. They believed that the two twins remains with the same age at their meeting after the trip of the traveler.

The most known is Dingle but I begin with Milne and Whitrow (1949) who wrote two articles to show that the two observers must have the same age when the meet again after the trip of one of them. The authors present two arguments. One is based on the concept of clock. They ask if "all natural processes which run naturally of themselves independently of what we may do may equally serve as clocks and give the same results." It is well known that for Einstein and after him the physicist community, the answer is "yes". However for Milne and Whitrow the answer is "no" and they base themselves on the concept of "physiological time" which is not equivalent to the "physical time". The new notion in physics of "physiological time" is presented without definition but for them it is enough to exclude the age difference. The second argument is based upon the symmetry between the two observers. This seems analogous to that presented above by Kracklauer and P.T.Kracklauer and Fischer. But Milne and Whitrow add that at the meeting they will have the same age. I feel that the reason of their opposition to the age difference is given by a combination of their two arguments. However it is not very clear how they connect them. I am not aware of reactions of other physicists about the opinion of Milne and Whitrow.

Dingle is the most notorious opponent to the theory of Relativity. It is also known as a vehement polemicist. In the thirties of the twentieth century took place a controversy about Cosmology and among the actors were Milne and Dingle (but not on the same side). People took him very seriously because he was a well-known physicist and philosophe of Science. Dingle remains a mystery because it is very difficult (if not impossible) to follow the evolution of his mind. He was a physicist who, as numerous physicists, accepted the Theory of the Relativity: he wrote two books on the Relativity: *Relativity for All* in 1922 and *The Special Theory of Relativity* in 1940.

However, in 1956 he published in *Nature* a paper in which he claims that at their meeting the twins have the same age. He does not claim that the theory is false but since "something must have happened to one [clock] which did not happen to the other and nothing is in question except their relative motion. Hence they cannot show different times on reunion". He writes also: "When the observers meet again they are both at the same place and relatively at rest. Their judgments of simultaneity agree, and so their clocks agree".

In the same volume of *Nature* an answer by McCrea is published. Two arguments are presented to oppose Dingle. First the traveler has a different history because he has a motor engine which permits him to leave the earth, to turn around the star and to come back. Secondly, the word-line of the stay at home is a geodesic but not that of the traveler.

Of course, there is also in the same volume of *Nature* an answer of Dingle. It concerns some details of his article and about the argument of the traveler with a motor engine, he eliminates it in saying that the accelerations are applied in so short times comparatively to the whole trip that they are unimportant.

We have to leave Dingle who continues during several years the controversy because he had now a new claim, namely that the complete theory of the Special Relativity is false. Scientists of History of Physics (see the detailed article of Chang) have difficulties in understanding the intellectual evolution of Dingle. He began his career as a "normal" physicist, had some doubts about the twin paradox and finished by a complete deny of the Special Relativity.

Is it necessary to try to understand the position of Dingle in his 1956 paper? I think that the answer is yes because he is one example (and there are many) of ignorance among physicists and scientists. He does not know the 1918 paper of Einstein, the 1934 Tolman's book and the 1952 Møller's book. Furthermore he made confusion about the two versions of the paradox. In the non-accelerated version the proper times of the twins are identical i.e. they have the same age and he is right. However, even if the acceleration times are very short, the problem remains the true story as shown above. And he is wrong since, even with a very short acceleration time, there is asymmetric aging.

In 1971, Sachs who was professor of physics at the State University of New York published in *Physics Today* (September 1971) an article in which he defends the view that at their reunion, the twins will have the same age. His first argument is to claim that "[the] correlation between a fixed observer time measure on a moving clock and its aging process is not necessarily true". For him "the space-time parameters are a useful language to *express* the laws of nature, whereas the physical interactions *obey* the laws of nature". In other words, when an observer measures the time of a process in a frame moving relatively to him, there is no one to one correspondence between the measuring time and with the proper time of the process in the moving frame. His reasoning goes as follows. a) there is no complete solution of the twin paradox even for those who accept the one to one correspondence mentioned in the preceding sentence. He criticizes the Tolman solution in that it is correct only for small velocities, (in what he is right) but he ignores the Møller solution. b) I quote from his article: "It appears to me that two identical clocks, synchronized in a single proper frame of reference, should be synchronized in all future times of observation, even though they may not appear to be synchronized unless they are again in the same proper frame of reference". c) a mathematical demonstration is presented in order to show that the variation of the proper time does not depend on the path in the space-time.

His concept of "proper time" implies that two clocks, once synchronized in one reference frame will be always synchronized in this frame whatever the story of each clock. For him it is the analogous of the case two identical rods which have their proper length in the same frame and when one is put in motion and comes back at rest in the original frame, they have again their proper length. In fact as shown above, the situation is more complex and the encounter of two clocks is not a simple phenomenon. This was already stressed by Einstein long time ago (see in the article of Pessic an interesting discussion between Einstein and one student). When the clocks are synchronized in the same frame they have the same rate, which is different (relatively one to another) if one is put in motion. If it comes back to the first frame, they have again the same rate but do not indicate the same time.

For Sachs the point b is essential and this means that he does consider possible changes due to the stopping of a frame. Sachs is consistent with all the textbooks on Special Relativity; no one considers the stopping of a frame. The time dilatation is only a phenomenon of frames in relative motion and Special Relativity is viewed as a world of non-stopping frames.

My trouble comes from the numerous answers to the paper of Sachs (*Physics Today,* January 1972). They do not convince Sachs because they do not consider the first two arguments evocated above. Only the mathematical proof is really disputed and the different authors bring several arguments that were surely well known to Sachs. (I cannot discuss the mathematical aspect of the Sachs paper since he uses quaternions in which I am not familiar). In fact two answers would be enough. The first is: "please, read the book of Møller" and the second "please, let us wait the results of the experience of J.C.Hafele and R.E.Keating". It was already known that Hafele and Keating were making a critical experience (Sachs himself quotes Hafele). Why these simple and sufficient answers were not given to Sachs? I feel that this is also a part of the phenomenon.

II 4 The mystery

Modern Physics has this so peculiar property that we must leave all intuitive notions if we intend to understand the physical world. The classic example is the behavior of particles in Quantum Mechanics. In some circumstances an electron may be a simple particle with a mass, a velocity, a position and a trajectory as we can imagine for a very small ball. But in other circumstances, the electron behaves as a wave defined by a field and exhibits the interference phenomenon and all the concepts associated with a "particle" must be left out. The physicist must live with this intuitive contradiction and in general he succeeds. However, Relativity is more demanding because it concerns time.

First one has to admit that a temporal phenomenon measured in a given reference frame A as $\tau_A$ is not observed with this value in all the other reference frames moving with a constant velocity relatively to A. In all these other frames the phenomenon is measured as $\tau > \tau_A$. However, if one considers the same temporal phenomenon in two reference frames A and B, the observers in each frame will measure the same time, what it is called the proper time.

Until this point, there is not too much difficulty. That the measure of the time depends on the reference frame is not too difficult to accept. This can seem strange but the experiment with the life time of mesons helps to accept the time dilatation from frame to frame.

But the final seems really absurd. If the frame B stops its motion (it has a special motor engine for that), the two clocks of A and B will not coincide even if they were once synchronized, the clock B will be behind the clock A. This begins to be too much. It seems unbelievable.

Of course, it is possible to accept all these conclusions but, at least for some people, this is accompanied by feeling of discomfort and even of anxiety. That an electron is sometimes a particle and sometimes a wave does not perturb my thought concerning my sensations and my life. But the fact that travelling will change time is difficult to stand. Maybe one can add the feeling of fear because all the basic intuitions concerning time are disputed. Even it is only a theoretical feeling; one comes to the conclusion that our basic instincts are no more sure and useful.

It was also possible to escape to this situation in following a proposal of Dingle (1964). It is known that, when one wants to solve a problem with the help of Mathematics, sometimes there are several solutions. It is habitual to exclude some of them because they have no physical meaning. Dingle proposed to do the same with Relativity. But now it is not possible: the experience of Kafele and Keating showed that the relativistic solution has an experimental basis. Here the anxiety and the fear increase because there is no simple explanation for this phenomenon.

My claim is that these feelings of anxiety and fear are very likely present among a large part of the physicists. However there are also other feelings. It is the pride to bring new vision of the world that philosophy cannot afford. Although it is not the place to discuss this point, I want only mention the tension which exists between physicists and philosophers. This also provides explanation of the scornful attitude of some physicists. One the most clear evidence is given by some answers to the article of Sachs in *Physics Today*. One can detect easily the annoyance[2] of those who believe to understand in front of the people which do not understand anything. But sometimes I ask myself if really people understand the problem[3] and its solution!

Here I have to open a parenthesis about Special Relativity. When Einstein in his fundamental 1905 paper describes the travelling of a clock A around a closed path and the meeting with a fixed clock B at the end of the path, he writes that the travelling clock A will be behind the fixed clock B. But what does he mean? In the context of his paper and of all the textbooks on Special Relativity, one can understand that the travelling clock A does not stop and observers associated with the clock B will measure a longer time for the travel than that the clock A. There is some ambiguity because we do not know what Einstein really intends: the clock A will stop or not? If it stops, it keeps its reading behind the clock B. One finds this ambiguity very frequently especially in textbooks.

As mentioned above, the large majority of the proposed solutions are in the frame of Special Relativity but excluding the Møller solution, in particular in textbooks. The problem was falsified a long time ago since a dynamical problem is transformed in a kinematical problem. By ignoring the Møller solution (voluntary or not voluntary) one remains in the ambiguity of the Einstein paper. However, people speak about departure and final stop but in the calculations it never shown that there is a final stop. The textbooks ignoring what happens at the stopping cannot give a correct solution and are obliged to give complicated solutions. In adopting such positions those who want to find the solution do not have the required tools for that. Consequently, they are obliged to invent several artifices like the exchange of signals.

---

[2] For example, D.A.Ljung writes in his answer to Sachs: "Correctly understanding the twin paradox is really just being careful in the use of special relativity.". In other words, his message can be read as that: "Poor Mr Sachs, you do not know how to use special relativity".
[3] Always among the people answering to Sachs one finds this sentence (V.Korenman): "..how many times must it be said that the *felt* acceleration of the travelling twin is sufficient asymmetry to remove the paradox". There is confusion between acceleration and force!

The desired goal is to find the famous ratio $\gamma$ of the two ages. How it is possible to show that the traveler is younger even when they do not meet? Consider the table 2 above in which the results are given in the case where one accepts the length contraction for the traveler (what I consider as not correct, see above). One remarks the agreement of the two observers: the times are identical. In the two frames, the time of the stay at home is $2L/V$ and the time of the traveler is $2L/\gamma V$. Since they agree about the times, it is as they are in the same frame.

But now there is a new paradox: one has two inertial reference frames moving with a velocity V relatively one to another and nevertheless there is agreement about the times. Relativity disappeared! By this way the fear associated with the stopping of the traveler is removed. Although the two observers do not have the same age, it is merely time dilatation like in the experiment of the meson decay. I think that by this way the twin paradox appears to be more bearable. I can add that may be to propose a new solution is a mean for being reassured that time dilatation is effectively a real phenomenon. It seems also that some need several different solutions to be convinced!

## II 5 Why Twins?

As mentioned above, although the idea of twins was given by Weyl 1922, it is only around the fifties of the preceding century that the name twin paradox was very often used. To find the hidden reason for this habit, I propose to investigate the collective image of twinship as it was elaborated trough myths and religions and to relate it to the use of the word.

I begin by quoting B.Beit_Hallahmi and M.Paluszny (1974) from their paper on twinship: "…there are common psychological elements in both mythological and scientific approaches to twinship. The major elements are fascination and ambivalence. Fascination with twin births has always been combined with a great deal of apprehension and ambivalence. In both primitive and modern societies, multiple births have been viewed as a potential source of familial and social conflict and complication."

Effectively one finds in the Bible and several mythologies frequent situations of conflict between twins. In the Bible one can mention Cain and Abel with the murder of Abel and also Jacob and Esau. Esau wants to kill his twin brother Jacob but Jacob has time to escape. In the Greek mythology (See the works of Aeschylus (1966)), the fight between Atreus (preferred by Zeus) and Thyestes (preferred by the people) with all the successive conflicts in the Atrides family is well known. In the roman mythology, the two twins Remus and Romulus (see the details of the story in Plutarch) who are the founders of Rome, are fighting for the government of the city. It results that Romulus kills his twin Remus.

In a complete different context, Levi-Strauss (1995) analyzed the myths of some North America and South America Indians and considers the antagonism between twins as a source of disequilibrium.

In fact, the twins' conflict is not completely general since there are also some cases where the twins manage their life in peace. For example in the Greek mythology the twins Castor and Pollux are in good relationship and there are taken as example of a peaceful twinship. Nevertheless the collective image of twins is frequently related to conflict and violence. This double presence of two identical persons is seen as something scandalous and abnormal. One of them must disappear or at least put in a bad position.

I propose to adopt this point of view considering the twins of the paradox. In the unconscious mind, the twinship must be destroyed and in the paradox this is what happened: one of the twins remains young and the other old, when young is better than old. I propose to interpret the habit to give the name "twin paradox" to the clock paradox as an intrusion of the subconscious in the language of physicists. All the efforts of the physicists are to show the asymmetry in the twins as it must be.

Conclusion

I recall that for Langevin and Einstein there is no paradox but only surprising results. When one considers the true and complete story, effectively there is no paradox since it is evident that the two observers are not in symmetrical positions. The traveler begins his trip applying a force and stops always by applying another force. The stay at home may calculate the travel time using the relativistic dynamics and the travel time of the traveler is obtained by calculating his proper time. It is so simple and so direct!

However, at an unknown time, the dynamical problem was transformed in a kinematical one. It seems that it was already the case in 1918 as it was presented by the critic in the 1918 Einstein paper. Einstein gives answers to an imaginary critic (may be not so imaginary, who knows?) who presents objections to the theory of Special Relativity.

The complete story was forgotten in favor of the kinematical aspect. The force that it is necessary to apply disappears and the result is the paradox. I suppose that people preferred an apparently simpler problem than a complicated one. However, it was never seen that it is a new and consequently a different problem. Moreover to find the "good" solution, one has to admit that the distance between Earth and Star is smaller for the traveler. Unfortunately this is not correct but for those who want to find the famous γ ratio, this is not very important! Complicated solutions or incomplete solutions are proposed but now, one has the "good" result.

I suspect that the uneasiness to accept all the consequences of the Special Relativity may explain why people prefer a problem without stopping (as in the kinematical version). The different ages of clocks after traveling and stopping contradict so violently our daily experience that people prefer not to see it. It is remarkable than in many solutions without acceleration, one speaks about departure and arrival but of course there are no departure nor arrival (since there are not acceleration), they are only words. The refusal to consider acceleration is not an innocent never a cleaver operation; it is a complete change of the problem and a refusal to see it in all its complexity.

I am sure that the twin paradox will provoke again and again papers since it is not a problem to solve but it is in itself a social and psychological phenomenon. There is no end.

# REFRENCES

H.Arzeliès  *Relativistic Kinematics (1966)* Pergamon Press, Oxford

D.Beit-Hallahmi  and M.Paluszy *Twinship  in Mythology and Science: Ambivalence, Differentiation,  and the Magic Bond* Comprehensive Psychiatry **15,** (1974) 345 - 353

D.Bohm   *The Special Theory of Relativity* (1965) Benjamin, New York

H.Chang *A Misunderstood Rebellion,  The Twin Paradox Controversy and Herbert Dingle's Vision of Science*   Stud.Hist.Phil.Sci. **24** (1993) 741 - 790

W.Cochran  *The Clock Paradox* Vistas in Astronomy **3** (1960) 78-87

C.G.Darwin *The Clock Paradox in Relativity*   Nature **180** (1957) 978-979

H,Dingle  *Relativity for all*

H.Dingle  *Special Theory of Relativity*

H.Dingle *Relativity and Space Travel*   Nature **177** (1956) 782-784

H.Dingle  Nature **177** (1956) 785

H:Dingle  *Reason and Experiment in Relation to the Special Relativity Theory*  Brit.J.Phil.Science **15** (1964) 41-61

R.Dugas *Sur les pseudo-paradoxes de la  relativité restreinte* Comptes Rendus Acad.Sci. **238** (1954) 49

A.Einstein   *Die Naturwissenschaften* **6**, (1918) 697.  *Published in The Collected Papers of Albert Einstein*  (1989) Princeton University Press, Princeton

E.Fischer  *The Myth of the Twin Paradox* (2010) arXiv:1008.0174v1[physics.gen-ph]

G.F.Fitzgerald   *The Ether and the Earth's Atmosphere*  Science **13** (1889)  390

A.P.French  *Special Relativity* (1968)  Norton, New-York

 R.M.Frye and V.M. Brigham   *Paradox of the Twins*   Am.J.Phys. **25** (1957) 553-555

Ø. Grøn  *The Twin Paradox in the Theory of Relativity*   Eur.J.Phys. **27** (2006) 885-889

A.Grünbaum *The Clock Paradox in the Special Theory of Relativity* Phil.Science **21** (1954) 249 – 253

E.L.Hill  *The Relativistic Clock Problem*  Phys.Rev **72** (1947) 236 – 240

J.C.Hafele and R.E.Keating  *Around-the-World Atomic Clocks: Predicted Relativistic Time Gains* Science **177** (1972) 166-168

J.C.Hafele and R.E.Keating *Around-the-World Atomic Clocks: Observed Relativistic Time Gains*  Science **177** (1972) 168-170

A.F. Kracklauer and P.T.Kracklauer  *On the Twin Non-paradox* (2000) arXiv:physics/0012041v1

P. Langevin  *L'Evolution de l'espace et du temps* Scientia **10** (1911), 31-34

Online fr.wikisource.org  (in French)

     cn.wikisource.org  (in English)

C.B.Leffert and T.M.Donahue *Clock Paradox and the Physics of Discontinuous Gravitational Fields*  Am.J.Phys. **26** (1958) 515 – 523

C.Levi-Strauss  *The Story of Lynx* (1995) University of Chicago Press, Chicago

H.Lichtenegger and L.Iorio *The Twin Paradox and Mach's Principle* (2011) arXiv:0910.1992v2

H.Lorentz in  A.Einstein, H.Lorentz, H.Weyl, and H.Minkovski, *The Principle of Relativity* (1923)  Dover Publications, New-York

W.M. McCrea  *Answer to Dingle*  Nature **177** (1956) 784-785

E.A.Milne and G.J.Whitrow *On the so-called "Clock paradox" of Special Relativity* Phil.Magazine  **40** (1949)  1244 - 1249

C.Møller  *The Theory of Relativity* (1952) Clarendon Press, Oxford

R.A.Müller *The Twin Paradox in Special Relativity* Am.J.Phys. **40** (1972) 966 - 969

T.Müller, A.King and D.Adis  *A Trip to the End of the Universe and the Twin Paradox* Am.J.Phys **76** (2008) 360 - 368

R.Perrin *Twin Paradox: a Complete Treatment from the Point of View of Each Twin* Am.J.Phys **47** (1979) 317 - 319

P.Pesic  *Einstein and the Twin Paradox* Eur.J.Phys. **24** (2003) 585-590

R.Resnick and D.Halliday *Basic Concepts in Relativity and Early Quantum Theory* (1972) John Wiley, New York

H. Reichenbach  *The Philosophy of Space and Time* (1958) Dover Publications, New York

W.Rindler *Relativity: Special, General and Cosmological* (2001) Oxford University Press, Oxford

R.H.Romer *Twin Paradox in Special Relativity* Am.J.Phys (1958) **27,** 131- 135

W.G.V.Rosser *An Introduction of the Theory of Relativity* (1964) Butterworths, London

M.Sachs *A Resolution of the Clock Paradox* Physics Today, September 1971, p.23

D.Styer *How Do Two Moving Clocks Fall Out of Sync? A Tale of Trucks, Threads, and Twins* Am.J.Phys. **75** (2007) 805 – 814

G.Thomson and WG Headlam *Aeschylus* (1966) Cambridge Univ.Press, Cambridgw

R.C.Tolman *Relativity, Thermodynamics and Cosmology* (1934) Clarendon Press, Oxford

H.Weyl *Space-Time-Matter* (1922) Methuen, London

Ta-You Wu and Y.C.Lee *The Clock Paradox in the Relativity Theory* Int.J.Theor.Physics (1972) **5**, 307-323